\documentstyle[12pt]{article}

\def\tg{$t_{2g}\/$\,} 
\def\eg{$e_{g}\/$\,}

\def\bil{La$_{2-2x}\/$Sr$_{1+2x}\/$Mn$_{2}$O$_{7}\/$\,}
\def\thb2{\frac{\theta_{i}}{2}}  
\def\Cp{C$^{\prime}$} 

\begin{document}

\begin{center}
{\Large {\bf Magnetic and orbital order in overdoped bilayer manganites}}\\
\vspace{1.0cm} 
{{\bf Tulika Maitra}$^{*}$\footnote{email: tulika@mpipks-dresden.mpg.de}  
and {\bf A. Taraphder}$^{\dagger}$\footnote{email: arghya@phy.iitkgp.ernet.in, 
arghya@cts.iitkgp.ernet.in}}\\ 

\noindent $^{*}$Max-Planck-Institut f\"ur Physik Komplexer Systeme \\
 N\"othnitzer Str. 38 01187 Dresden, Germany 
\vspace{0.3cm} 

$^{\dagger}$Department of Physics \& Meteorology and  
Centre for Theoretical Studies, \\
Indian Institute of Technology, Kharagpur 721302 India \\   
and\\
Abdus Salam International Centre for Theoretical Physics \\
Strada Costiera 11, 34014 Trieste, Italy. 
\end{center}  

\begin{abstract}
\vspace{.5cm} 

The magnetic and orbital orders for the bilayer manganites in the doping 
region $0.5 < x <1.0$ have been investigated from a model that incorporates 
the two \eg orbitals at each Mn site, the inter-orbital Coulomb interaction 
and lattice distortions. The usual double exchange operates via the 
\eg orbitals. It is shown that such a model reproduces much of the 
phase diagram recently obtained for the bilayer systems in this range 
of doping. 
The C-type phase with ($\pi,0,\pi$) spin order seen by Ling et al. appears
as a natural consequence of the layered geometry and is stabilised by the
static distortions of the system. 
The orbital order is shown to drive the magnetic order while the anisotropic
hopping across the \eg orbitals, layered nature of the underlying structure and 
associated static distortions largely determine the orbital arrangements. 

\end{abstract}
 
\noindent PACS Nos. 75.47.Gk, 75.30.Et 
\vspace{.5cm} 

It has been realised \cite{khom,akimoto,tmat} in the recent past that 
the physics of the region $x > 0.5$ is quite different from that in 
the $x < 0.5$ for the 3D manganites and one has to look at the heavily 
doped ($x > 0.5$) manganites from a different perspective. 
A similar situation prevails \cite{ling} in the bilayer manganites, 
the $n=2$ member of the Ruddelsden-Popper series $(R,A)_{n+1}Mn_{n}O_{3n+1}$ 
(where $R$ and $A$ are rare-earth and alkaline-earth ions respectively)
as well. The doping region $0 < x < 0.5$ for bilayer manganites has 
been investigated in some detail and a rich variety of phases identified.
These layered systems also show large magnetoresistance (MR) and a sequence 
of magnetic phases \cite{kimura,hirota} like their 3D counterparts. 
From a ferromagnetic (FM) state at low doping ($x\simeq 0 $) to canted
antiferromagnetic (AFM) metallic to AFM insulating state between $x=0.37$ 
to $x=0.48$ have been reported \cite{hirota,kubota,sury}. At $x=0.5$
there is a possible coexistence between charge ordered (CE-type) and
layered A-type spin ordered state \cite{kubota,cold}.  

The region $x > 0.5$ has now been investigated \cite{ling,billinge} carefully
using neutron 
scattering and a succession of magnetic phases A$\rightarrow$ C$\rightarrow$ 
G has been observed. Between the A- and C-type phases (between $0.66<x<0.74$), 
there appears
a region of no well-defined long range order (LRO). Beyond $x > 0.74$
the AF C-type spin order is seen (along with a polytype, where the long
c-axis is doubled). Interestingly, in the C-phase (or its polytype), the
spins are aligned in the long basal plane b-axis, along which there
is a distortion concomitant at $x=0.74$.
In addition, both A-type and C-type phases have been found to be orbitally 
ordered. There is no evidence of canting of spins in the region $x > 0.5$. 

The role of orbitals on the underlying magnetic order is stressed 
\cite{akimoto} already in the context of the various magnetic structures 
of the 3D manganites. Models have been proposed \cite{khom,tmat} 
for the manganites that incorporate the $e_g$ orbitals and the anisotropic 
hopping between them. It was also realized that the inter-orbital 
interaction is quite crucial for the underlying orbital order 
\cite{tmat,hotta}. The use of such models to the bilayer manganites 
(like \bil) has only had limited success though \cite{pai,maezono}. The 
quasi two-dimensional nature of 
the underlying lattice stabilises the A-type layered magnetic structure and 
the models have not been able to reproduce the observed C-type 
one-dimensional 
magnetic structure. The A-type AFM instability is indeed quite strong
in the layered system (see fig. 1 in Ling et al \cite{ling}), extending 
from $x=0.42$ to 0.66. Moreover, at low temperatures, the CE-type spin 
and charge order seems to be absent and replaced by the A-type spin 
order \cite{moritomo},
even at $x=0.5$. On the other hand, there is a tetragonal to orthorhombic 
transition (elongation of the basal plane b-axis \cite{ling,billinge}) near 
$x=0.74$ where the C-phase appears. There is no buckling of
the octahedra associated with these distortions.  
The nature of spin and orbital ordering, as suggested by Ling et al.
\cite{ling} and Qiu et al. \cite{billinge}, clearly points to the role of 
the electron-lattice coupling and 
the resulting elongation of the b-axis on the magnetic and orbital structure. 
Both the A- and C-phases are orbitally ordered and there is intimate
connection between the preferred orbital orders, the lattice distortions
and the magnetic order.

The experimental observations and theoretical understanding generated for
the heavily hole-doped 3D manganites quite naturally lead to a model 
for the bilayer manganites in the region of doping $x > 0.5$ . The model 
incorporates the degenerate \eg manifold and the physics of double exchange 
(DE) along with electron-electron and electron-lattice interactions. 
Such a model is given by  

$$ H = J_{AF} \sum_{<ij>} {\bf S_{i}}.{\bf S_{j}} - J_{H} \sum_{i} {\bf S_{i}}.
{\bf s_{i}} - \sum_{<ij>\sigma,\alpha,\beta}t_{i,j}^{\alpha \beta} 
c_{i,\alpha,\sigma}^{\dagger} c_{j,\beta,\sigma} + H_{int}+H_{e-l}\eqno(1)$$

Here ${\bf S_{i}}$ and ${\bf s_i}$ represent the \tg and \eg spins at 
site $i$ and $J_H$
and $J_{AF}$ are the Hund and super-exchange (SE) coupling respectively. 
The usual charge and spin dynamics of the conventional DE model operate 
here too, with additional degrees of freedom coming from the degenerate \eg 
orbitals ($\alpha, \beta$ take values 1 and 2 for the two \eg orbitals). The 
hopping across them is determined by the symmetry of $e_g$ orbitals. 
The term $H_{int}= U^{\prime} \sum_{i\sigma\sigma^{\prime}}\hat{n}_{i1\sigma}
\hat{n}_{i2\sigma^{\prime}}$ describes on-site inter-orbital interaction. The 
intra-orbital term does not play a significant role for the typical values 
of $J_H$ one is working with \cite{hotta,misra}. The inter-bilayer exchange
interaction is known to be at least a 100 times weaker \cite{fujioka} than the 
intra-bilayer one. Two bilayers are also well-separated in an unit cell and
intervened by the rare-earth ions. This allows us to consider only one bilayer 
for the calculations that follow. 

At $x=1$ the $e_g$ band has no electrons and the physics is governed 
entirely by the AF superexchange between the neighbouring $t_{2g}$ 
spins. On doping, the band begins to fill up (with nominal electron-density 
$\frac{1-x}{4}$). In the absence of electron-lattice coupling, the 
kinetic energy (KE) of electrons in the $e_g$ band along with the attendant 
Hund's coupling between $t_{2g}$ and $e_g$ spins begin to compete with the 
antiferromagnetic SE interaction leading to a rich variety of magnetic and 
orbital structures. The JT distortions, through the local electrostatic 
coupling (acting as an `orbital magnetic field'), lift the degeneracy 
of the \eg orbitals and affect the DE mechanism considerably.  

The coupling between the \eg manifold and lattice is incorporated through
a term in $H$ \cite{kanamori}, 
$$H_{e-l} = g \sum_{i,m}\tau_{i,m} {\bf Q}_{i,m}$$
\noindent where $Q_{i,m} (m=1,2)$ are the even-parity local distortions of an 
MnO$_{6}$ octahedron and $\tau_{1}$ and $\tau_{2}$ are the 
first and third Pauli matrices. A positive sign of $g$ renders the 
$3z^{2}-r^{2}$ orbital stable over $x^2-y^2$ orbital for $Q_3$ distortion as 
there is negative charge on the surrounding oxygen ions. 

Writing $Q_{i,1}=r_{i}sin\theta_{i}$ and $Q_{i,2}=r_{i}cos\theta_i$, $H_{e-l}$ 
is diagonalised by the unitary transformation in the local \eg orbital space to 
$S_{i}H_{e-l}S_{i}^{-}$ where ${\bf S_i}= \left( \begin{array}{cc}
cos\thb2& sin\thb2\\ -sin\thb2 & cos\thb2 \end{array}\right)$.  
The choice of $\theta_i$ determines the orthogonal combination of orbitals 
and is dictated by the physics at hand. In addition, the orbital pseudospin 
operator turns out to be $<\vec{\tau_{i}}>=(sin\theta_i, 0, cos\theta_i)$. 
The hopping matrices $t_{\alpha,\beta}$ along $x, y, z$ directions, therefore, 
transform as $S_{i} t^{\hat{x},\hat{y},\hat{z}} S_{i}^{-}$. The rotational 
symmetry in the orbital space implies $H_{int}$ remains invariant. 

The diagonalisation of the KE part of $H$ leads to two bands. 
In the pure (uncanted) phases the bands in A- and C-phases become purely 
two- and one-dimensional. However, even in the presence of canting there 
is little dispersion along the AFM aligned directions - a plane in C-phase or 
a line in A-phase. Typical values of the interaction and band parameters for 
the bilayer systems are in the same range as in the 3D manganites. 
The Hund coupling and Coulomb correlations are the largest scale of 
energy \cite{misra,hotta} in the problem. Treating the t$_{2g}$ spins 
classically, the SE contribution to the ground state energy becomes 
$E_{SE}=\frac {J_{AF}S_0^2}{2}(2cos\theta_{xy}+cos\theta_z)$ where 
$\theta_{xy}$ and $\theta_{z}$ are the angle between the near-neighbour (nn)
\tg spins in the $xy$ plane and $z$ direction respectively.
 
For an uncanted homogeneous spin configuration in the ground state, we 
choose ${\bf S}_i={\bf S}_{0}\exp
(i{\bf q.r_i})$ where the choice of ${\bf q}$ determines different spin 
arrangements for the $t_{2g}$ spins \cite{tmat}. We begin our discussion 
by considering the model without the Coulomb interaction terms $U^\prime$. The 
nn Coulomb interaction and its effects will be dealt with later. 

Using the semi-classical approximation for the $t_{2g}$ spins the Hamiltonian 
(1) reduces to an $8\times 8$ matrix. The distortions are assumed to be 
uniform ($r_{i} = \sqrt{Q^{2}_{1}+Q^{2}_{3}}= r$). In almost 
all the manganites, the JT energy scales ($2|gr|$) are nearly in the same
order as the band-width, about 1$eV$. A typical value of $|gr|$ is therefore
about 0.5$eV$ at $x=0.55$ \cite{ftnt1}, where the tetragonal distortion is 
largest, Mn-Mn distance along c-direction shortest. The value of $gr$ 
gradually decreases with increasing $x$ as the c-axis elongates and 
vanishes by $x\simeq 0.9$. Around $x=0.75$ there is a tetragonal to 
orthorhombic transition, with slight elongation of the basal b-axis 
disappearing by about $x=.92$.  
It is argued \cite{billinge} that due to possible delocalisation of \eg
electrons, the self-consistent JT scale around $x=0.55$ could be much less.
On the other hand there is evidence of charge ordering close to this
region \cite{sury,cold,billinge}, which would lead to incipient localisation 
of charges. Nevertheless, the scale of static JT distortion used here 
is the bare value corresponding to an MnO$_6$ octahedron. 

We use mean-field approximation \cite{tmat,hotta,maezono} to treat the 
Hamiltonian. This is shown to work quite well for the ground state
properties \cite{hotta} in the 3D manganites. 
The mean-field Hamiltonian is diagonalised at each {\bf k}-point 
on a momentum grid. The ground state energy is calculated for different 
magnetic structures. We consider four different magnetic structures 
relevant for the experimental phase diagram. These are  (with q values in
the parentheses) A-type ($0,0,\pi$), C-type ($\pi,\pi,0$)-we call as C-type
the usual C-phase with FM chains along c-direction, C$^\prime$-type
($\pi,0,\pi$) and the 3D AFM G-type  ($\pi,\pi,\pi$). The third one is
the same as a C-type, only that its FM ordering is along b-direction as 
reported by Ling et al. The magnetic 
structure with minimum ground state energy is determined for each set of 
parameters ($x$, $J_H$, $J_{AF}$) for the range of doping ($0.5 < x \le 1$) 
for a given distortion. Fig. 1a shows the ground state energy (all energies
are measured in terms of $t^{\hat{z}}_{22}=t=0.25eV$) with doping
$0.5 < x < 1.0$ for typical values of exchange interactions for $|gr|=0$ 
and 2.0 (along the c-axis) and Fig. 1b shows the same with a distortion
along b-direction nearly half the magnitude. The energies for $|gr|=0$
are offset by 0.2 in order for better viewing. 

On shortening the bond lengths along c-axis, the energy of the C-phase 
rises while energies of both A and C$^\prime$ phases go down. A-phase with its
planar FM magnetic and orbital order (discussed below) is clearly 
favoured over the C-phase with out-of-plane FM magnetic (orbital) order. 
The C$^\prime$-phase, with FM spin order along b-direction, also gains from the 
contraction in c-direction. This is even more apparent in Fig. 1b where 
an elongation in basal b-direction stabilises C$^\prime$-phase further. 
As reported in previous work
\cite{pai,maezono,okamoto} A-phase instability is quite strong in
the layered manganites owing to the 2D structure of the DOS. The static 
distortion along b-direction stabilises both A and C$^\prime$ phases, while the
gain in stability of C$^\prime$ phase is larger than that of A primarily 
due to its 1D magnetic (and orbital) order along b-direction. 

The phases A and C$^\prime$ are both orbitally ordered. Shown in fig. 2, 
the A-phase
has planar $x^2-y^2$ order while the C$^\prime$ phase has $3y^2-r^2$ order. The
orbital densities do not change over continuously, there is an abrupt change
across the A-C$^\prime$ transition between the two sets of orthogonal orbitals 
indicating a first order transition between them. A strong orbital order 
is also seen \cite{hotta} in exact diagonalisation study. Although the 
staggered orbital order is favoured close to half-filling, the second
order $t^2/|gr|$ process is inoperative at this low electron-doped region 
where orbitals are mostly unoccupied. 

A phase diagram is then obtained in the $|gr|-x$ plane for typical values of
$J_HS_0$ and $J_{AF}S^{2}_{0}$. It is observed (Fig. 3a) that with increasing
$|gr|$ along c-direction, the C$^\prime$ state stabilises slightly. The 
GC$^\prime$ boundary 
is hardly affected as there are few electrons there. The large $x$ part of 
the phase diagram is similar to 3D manganites primarily due to the absence 
of any significant energy scales other than SE energy at such low 
electron-densities and reproduces the 3D AFM G-phase.
The effect of elongation of the b-axis is more prominent as discussed above.
The \Cp phase stabilises considerably over the A phase due to the changes
in the occupied \eg DOS with enhanced orbital ordering of $3y^2-r^2$. The 
effect of change of bond lengths and consequent enhancement in bare hopping 
may stabilise A-phase somewhat when the c-axis contracts. The 
elongation in b-direction can also reduce the hopping in that direction 
thereby reducing the stability of \Cp. The stabilisation coming from the static 
JT effects are expected to be stronger than changes coming from enhanced 
hopping at the doping regions considered. With changes in bond 
length less than 10\% \cite{ling,billinge}, and the density of electrons low,
this effect may not be large. In addition, the spin exchanges also depend on 
bond length (higher order in $t$ as $J \sim t^{2}$). Such effects are 
neglected in the presentation here. 

The phase diagrams in $J_HS_0 - x$ (Fig. 4) and $J_{AF}S^{2}_{0}-x$ (Fig. 5) 
reflect similar physics. To compare the theoretical phase diagram with
experiments, in Fig. 5a, we have included the actual distortions between 
$0.5 < x < 0.92$ with $|gr|=2.0$ at $x=0.55$ going down as $x$ increases
(by $x=0.75$ the lattice nearly relaxes in the c-direction) \cite{ling}.
The distortion in b-direction is smaller and occurs between $ 0.75<x<0.92$.  
The phase diagram resembles the experimental one, albeit without the
region of no spin order between $0.66 <x < 0.74$. The \Cp phase in Fig. 4a 
is more stable than that seen in experiments, covering this region of $x$ 
where no apparent LRO is seen. Although the model recovers the \Cp phase seen 
in experiments, rather than the large A-type region observed in previous work
\cite{pai,maezono,okamoto}, it overestimates the stability of this phase even 
without any static distortion. Note that there is a ferromagnetic phase
in fig. 5 at very low $J_{AF}$ where the DE mechanism dominates. 

Canting of the magnetic structures ${\bf S}_i$ is included via ${\bf S}_i=
S_0(sin\phi_i, 0, cos\phi_i)$ with $\phi_i$ taking all values between 0 
and $\pi$. In the G-phase, at large J$_H$, there is a small canting
in the $xy$-plane ($\sim 8^{o}$, inset in Fig. 4b), while $\theta_z$ does not 
cant. The physics 
is quite similar to the 3D manganites \cite{tmat} and the $x\sim 0$ region
of bilayer systems \cite{maezono}. At large J$_H$ in the G-phase the 
KE gain of the \eg electrons through DE, via the generation of an FM 
component of the underlying \tg spins, more than offsets the `cost'
of tilting \tg spins away from magnetically ideal AFM state. Tilting
in the $xy$-plane leads, of course, to a larger gain in KE than canting
in $\theta_z$, which remains insignificant. At smaller J$_H$ in the 
G-phase and in the A- and C$^\prime$-phases, this mechanism is energetically
inconsequential
and we do not find any canting which is also reflected in the discontinuous 
(1st. order) change in the orbital order across A-C$^\prime$ transition. 

We include the inter-orbital interaction term in the mean-field. 
As in 3D manganites \cite{tmat}, this term immediately stabilises \Cp phase. 
The 1D instability of \Cp state is more favourably affected by the 
inter-orbital interaction 
and preferential occupation of orbitals due to $U^\prime$. In addition, the
higher electron-density in the A-phase makes this phase vulnerable to
Coulomb interactions compared to the \Cp or G phase at lower electron-density
\cite{hotta}.

The entire
phase diagram with its magnetic and orbital order owes its origin to the
competition between DE mechanism, SE interaction, electron-lattice coupling
and electron-electron interaction. In the region $x \sim 0.5$, where the
electron-density is larger, the DE interaction via the degenerate \eg orbitals 
dominate. In the reduced dimensionality of the layered structure, the planar
$d_{x^2-y^2}$ orbital order along with DE coupling forces the ab-plane into
an FM configuration. The absence of long range correlation along c-direction 
and loss of tunnelling across the planes (driven by orbital order) induce 
AFM ordering in that direction and result in an A-phase. With a contraction 
of $MnO_6$ octahedra in the c-direction, this phase further stabilises. 
Without a coherent charge transport in the c-direction in bilayer systems, 
the C-phase with $(\pi,\pi,0)$ magnetic order is unfavourable in 
comparison to the A-phase as already observed 
\cite{pai,maezono}. Towards the $x=1$ end, where the \eg levels are empty,
the SE interaction brings about a $(\pi, \pi,\pi)$ magnetic order as
in the 3D case. The \Cp-phase, on the other hand, allows for coherent 
tunnelling in the b-direction, its 1D orbital order stabilises on contraction 
of the c-axis and elongation in the b-direction. At a certain $x$, as 
the electron-density reduces, this state stabilises over A-phase. The static 
JT distortions present in the system stabilises it until the SE interaction 
takes over at extreme low electron-doping. 
In the 3D manganites, the orbital order drives the magnetic order 
\cite{khom,akimoto} in the heavily hole-doped region. In the bilayer systems 
also, it is the orbital order, driven by the DE mechanism, anisotropic hopping 
across \eg orbitals and lattice distortions that seems to induce 
different magnetic phases.  

The scenario borne out here is markedly similar to the experimental
phase diagram and orbital order (fig. 13 in Ling et al.) in the bilayer
manganites. This also agrees
quite well with the observed phases in Qiu et al. The existence of a region
with no long range magnetic order around $x\sim 0.70$ is quite possibly
a result of the competing ground states with such close energies (fig. 1).
The A to \Cp transition being 1st. order in nature here there is a phase 
separated region (possibly dispersed due to long range Coulomb interactions).  
It would be interesting to look for inhomogeneous magnetic structures
\cite{billinge} or short range ordered phases (which are dispersed) in 
that region using more direct imaging techniques. It is also likely that
with longer range Coulomb interactions included, charge ordered regions
may stabilise close to $x=0.5$, seen in several experiments 
\cite{cold,billinge} recently.
 
\vspace{0.5cm} 

\noindent {\bf Acknowledgement} 
\vspace{0.5cm} 

We acknowledge useful discussions with S. K. Ghatak, S. D. Mahanti and 
G. V. Pai. TM acknowledges discussion with S. Wirth and AT acknowledges
some clarification of data by C. D. Ling. Research of AT was partly 
supported by DST (India) through Indo-US project.

\newpage
\center {\Large \bf Figure captions}
\vspace{0.5cm}
\begin{itemize}

\item[Fig. 1] Ground state energy of different magnetic phases versus hole 
concentration $x  > 0.5$ with and without lattice distortions. In (a) the
distrotion is in the c-direction while in (b) it is in the b-direction.
The $|gr|=0$ lines have been vertically offset by +0.2 to separate them 
from the lower bunch. 
 
\item[Fig. 2.] Orbital densities in (a) A- and (b) \Cp-phase at different 
values of parameters. In (a) the filled and open symbols are for $d_{x^2-y^2}$
and $d_{3z^2-r^2}$ orbitals. In (b), they represent, respectively, the
$d_{3y^2-r^2}$ and $d_{x^2-y^2}$ orbitals. The A- and C-phases are stable
only in part of the range of $x$ (see text). Note the sum of two orbital
densities is equal to $(1-x)/4$, the actual electron density. 

\item[Fig. 3.] Magnetic phase diagram in $|gr|-x$ plane. Note the gradual 
shrinking of the A-phase in the region  $x > 0.5$  while the G-phase remains 
nearly unaffected. 

\item[Fig. 4.] (a) Magnetic phase diagram in doping ($x$) - J$_{H}$S$_{0}$ 
plane is shown in solid line for experimentally relevant values of $|gr|$. 
In (b) is shown
the effect of $U^\prime$ on the phase diagram (at $|gr|=0$). The solid line 
is for $U^\prime=8$ and the dotted line in (a) and (b) are for $|gr|=
U^\prime=0$. In the inset in (b) is shown the canting of spins (away from
$\pi$) in the G-phase as a function of J$_{H}$S$_{0}$. 

\item[Fig. 5.] Magnetic phase diagram in doping ($x$) - J$_{AF}$S$^{2}_{0}$ 
plane. (a) and (b) correspond to similar situations as in Fig. 4 (a),(b). 

\end{itemize}
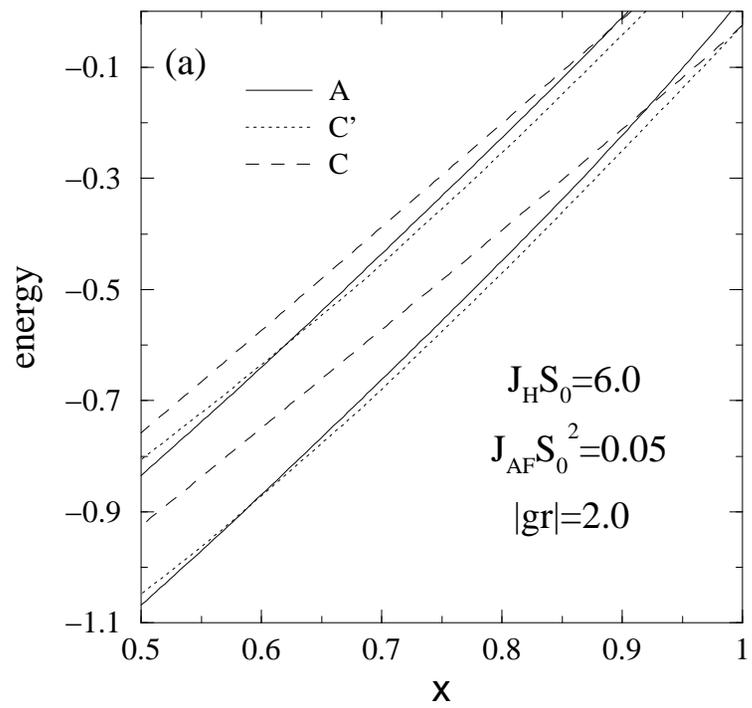
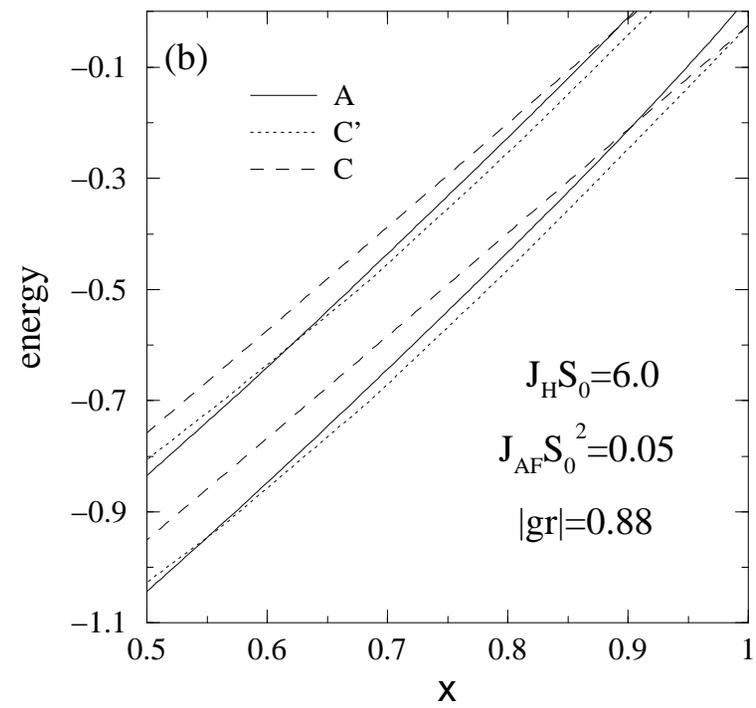
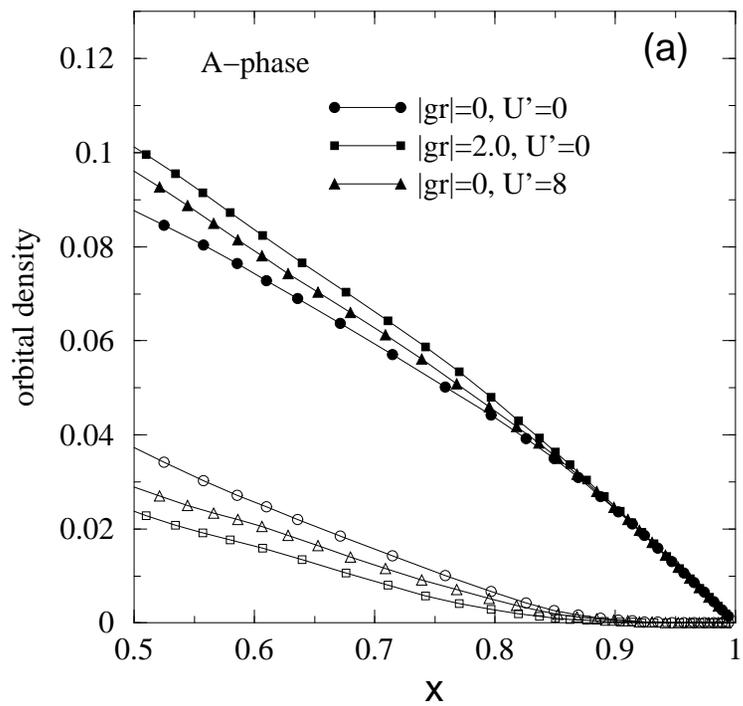
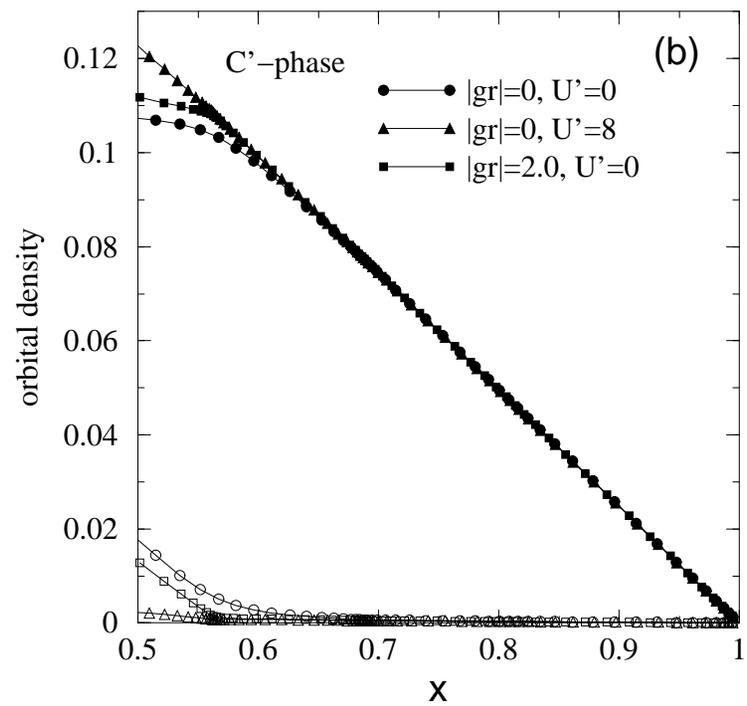
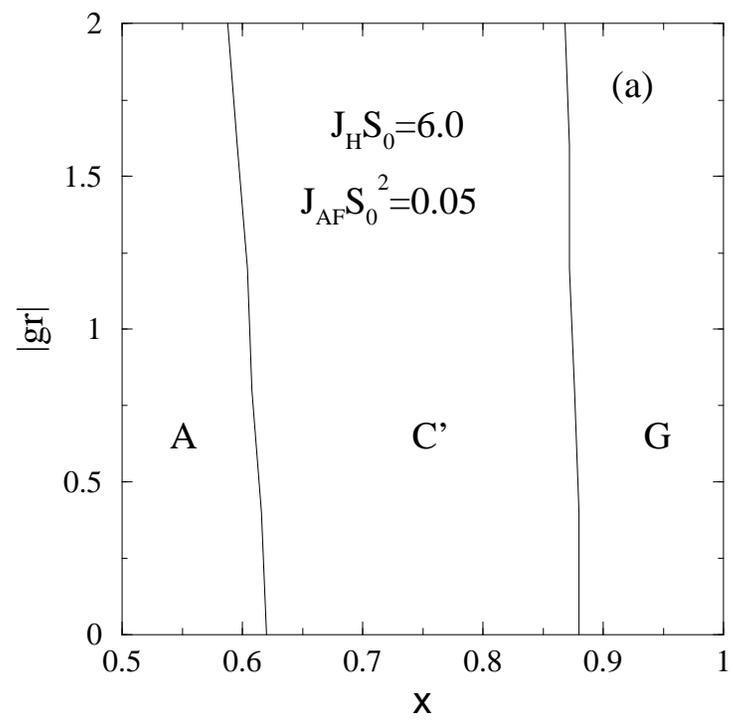
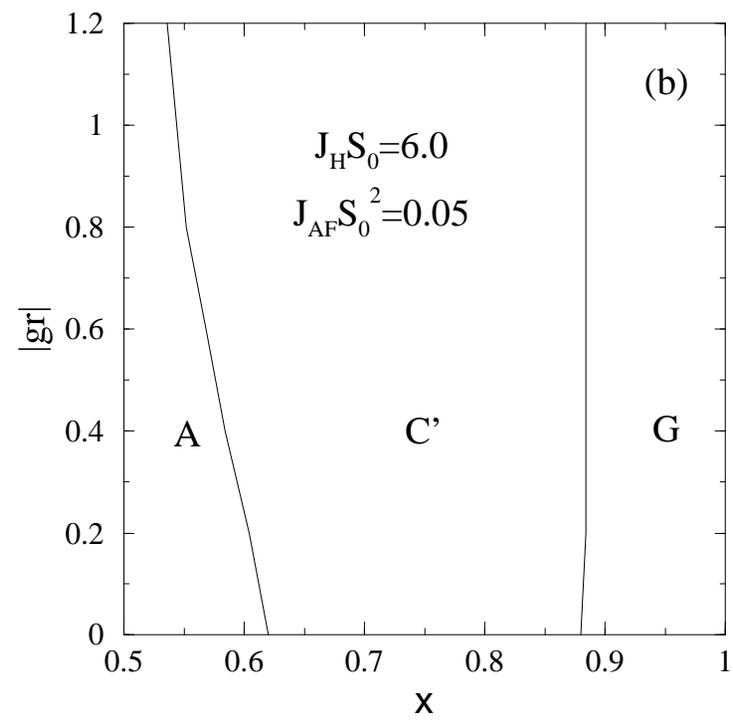
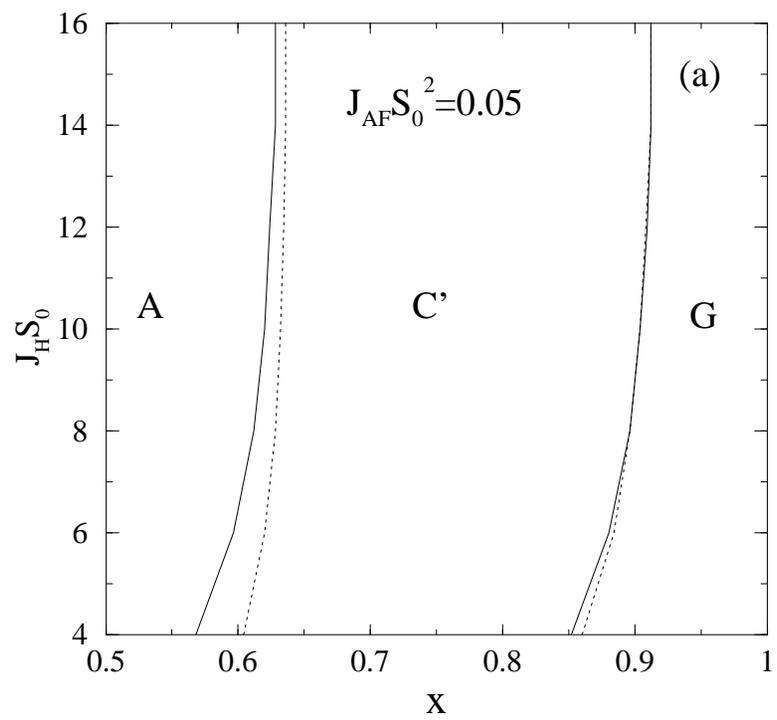
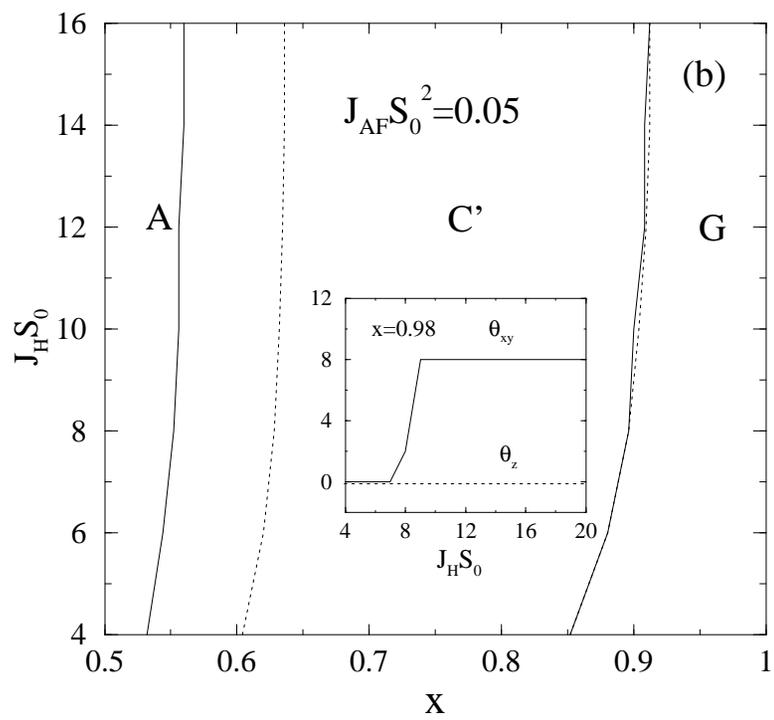
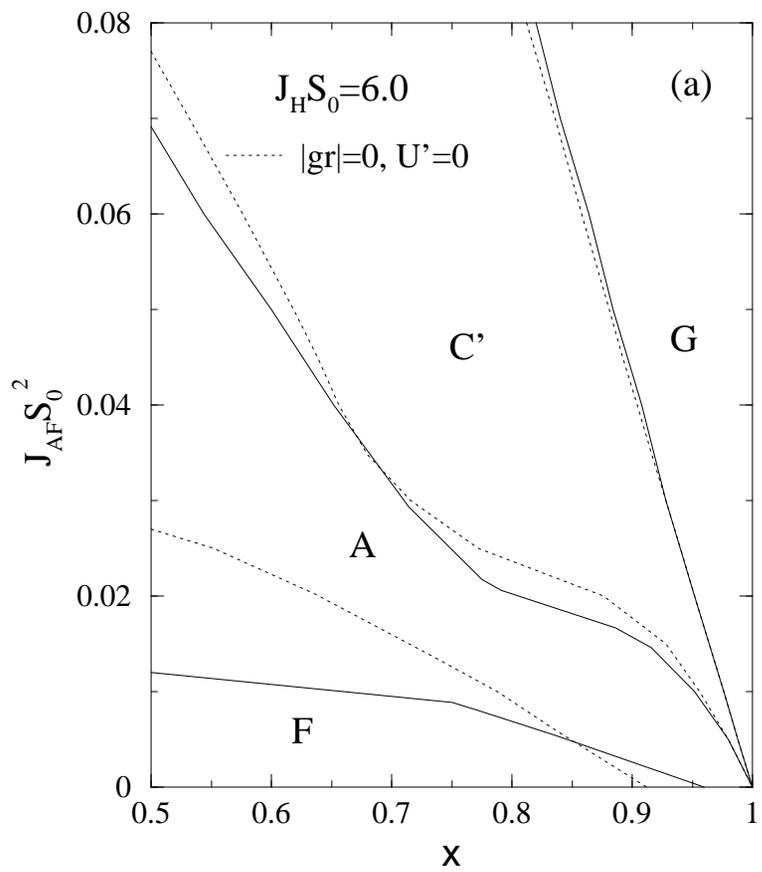
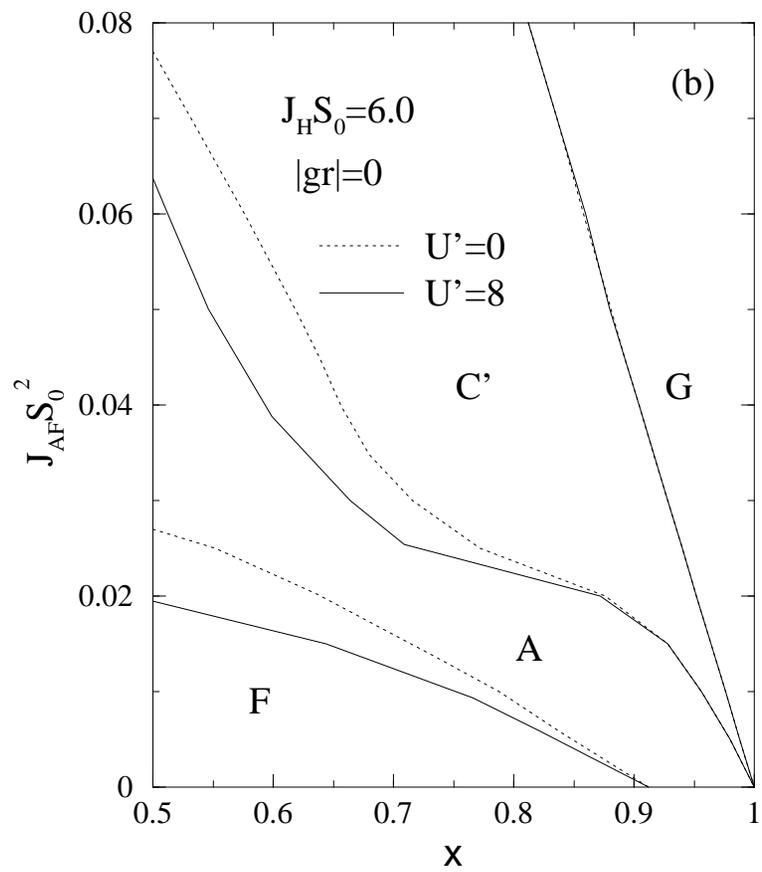
\end{document}